\documentclass[english]{article}
\usepackage[T1]{fontenc}
\usepackage[latin9]{inputenc}
\usepackage{esint}

\date{}

\DeclareRobustCommand{\greektext}{%
  \fontencoding{LGR}\selectfont\def\encodingdefault{LGR}}
\DeclareRobustCommand{\textgreek}[1]{\leavevmode{\greektext #1}}
\DeclareFontEncoding{LGR}{}{}
\DeclareTextSymbol{\~}{LGR}{126}

\makeatother

\usepackage{babel}
\begin{document}

\title{Entropy of isolated horizon from surface term of gravitational action}

\author{ Cheng-Yong Zhang$\ensuremath{^{1}}$%
\thanks{zhangcy@sjtu.edu.cn%
}, Yu Tian$\ensuremath{^{2,4}}$%
\thanks{ytian@ucas.ac.cn%
}, Xiao-Ning Wu$\ensuremath{^{3,4,5}}$%
\thanks{wuxn@amss.ac.cn%
}, Shao-Jun Zhang$\ensuremath{^{1}}$%
\thanks{sjzhang84@sjtu.edu.cn%
}}

\maketitle

{\centering\emph{
1. Department of Physics and Astronomy, Shanghai Jiao Tong University,
Shanghai 200240, China\\
2. School of Physics, University of Chinese Academy of Sciences,
Beijing 100049, China\\
3. Institute of Mathematics, Academy of Mathematics and System Science,
Chinese Academy of Sciences, Beijing 100190, China\\
4. State Key Laboratory of Theoretical Physics, Institute of Theoretical
Physics, Chinese Academy of Sciences, Beijing 100190\\
5. Hua Loo-Keng Key Laboratory of Mathematics, CAS, Beijing 100190,
China}}

\begin{abstract}
Starting from the surface term of gravitational action, one can construct
a Virasoro algebra with central extension, with which the horizon entropy
can be derived by using Cardy formula. This approach gives a new routine
to calculate and interpret the horizon entropy. In this paper, we
generalize this approach to a more general case, the isolated horizon,
which contains non-stationary spacetimes beyond stationary ones. By imposing appropriate boundary
conditions near the horizon, the full set of diffeomorphism is restricted
to a subset where the corresponding Noether charges form a Virasoro
algebra with central extension. Then by using the Cardy formula, we can derive the entropy
of the isolated horizon.
\end{abstract}

\section{Introduction}

Since the discovery of the Bekenstein-Hawking entropy of black holes \cite{Bekenstein,Hawking},
physicists have put enormous enthusiasm and endeavor to explain
its microscopic origin. One of the approaches is to use symmetry to
count states which was proposed by Strominger \cite{Strominger}. This
method can be traced back to Brown and Henneaux's work in 1986 \cite{Brown-Henneax}.
They found that, under the imposed boundary conditions at infinity, the asymptotic symmetry group of $AdS_{3}$ is a pair
of Virasoro algebra with central extension, implying that any consistent
quantum theory of gravity on $AdS_{3}$ is a conformal field theory
(CFT). While the Cardy formula \cite{Cardy-formula1,Cardy-Bloete,Cardy-formula2}
determines the asymptotic density of states of CFT entirely in terms
of the Virasoro algebra and is independent of other details of the
theory. So Strominger computed the black hole entropy by using the Cardy
formula and reproduced the standard Bekenstein-Hawking entropy of
Ba\~nados-Teitelboim-Zanelli (BTZ) black hole. Later, this method was developed
by Carlip \cite{Carlip1,Carlip2}. He proposed a set of boundary conditions
near the horizon rather than at infinity which also leads to a Virasoro subalgebra with a calculable
central charge. Several other related approaches have
been developed. One can refer to \cite{Mahji-Padm} and the references
therein for a review.

In all of the works mentioned above, the analysis is on shell and the bulk
action is employed. Motivated by the fact that York-Gibbons-Hawking
term \cite{York,Gibbons-Hawking} is also closely related to the entropy
of the horizon, and the surface and bulk terms of action encode the same
amount of information \cite{surface-Bulk}, Majhi and Padmanabhan (MP)
introduced Noether current associated with the surface term of the
gravity action \cite{Majhi-GH surface,Majhi-EH action}. The diffeomorphisms
related to the Noether current preserve the near horizon metric in
some non-singular coordinates. Given the Noether current and diffeomorphism,
there is a natural Virasoro algebra with central charge. The central charge and zero mode
eigenvalue of the Fourier modes of the charges then lead to the Bekenstein-Hawking
entropy via the Cardy formula. In \cite{zhang-wang}, this work
was generalized to the modified gravity with high curvature corrections
and the corresponding Wald entropy was derived. This approach is also successfully generalized
to other cases \cite{bigravity}. It shows that the approach based
on Virasoro algebra and central charge from the surface term of gravitational
action is general.

The spacetimes considered in the previous work are all static. In this paper, we want to extend this approach to more general spacetimes, non-stationary ones, which will not only cover static spacetimes, but also stationary and non-stationary ones. However, for general non-stationary spacetimes, event horizon is hard to define and is not applicable for calculation. Here the isolated horizon (IH) proposed by Ashtekar can be an
appropriate choice to realize our purpose \cite{IH_introduce}. Isolated horizon is a generalization
of Killing horizon of black hole. It is well known that all stationary
horizon satisfy the definition of isolated horizon \cite{Stationary-IH}.
For the non-stationary case, its near horizon spacetime can also be described by the isolated horizon  \cite{non-stationary-IH}.
In general, we do not have the explicit form of the metric on the IH from
its definition. Luckily, in MP's method, only the near horizon limit
of metric of IH \cite{NHLM-IH} is needed.
Thus we can calculate the entropy of IH based on the surface term.
This generalization is not straightforward since we only know the
near horizon limit metric of IHs. In previous work, the spacetimes
have well defined symmetries. For example, the spacetimes considered
in \cite{zhang-wang} are spherical symmetry. So they need only
to consider the invariance of the $r-t$ plane and the $r,t$ components
of the diffeomorphism vector. The other components vanish. Here, in our case,
we do not know the symmetries of the near horizon metric of IHs. There
is no special reason why the other components vanish. Thus we
should consider the general cases. Unlike the previous work, we impose
a suitable boundary conditions near the horizon from the very beginning to restrict
the asymptotic diffeomorphism vectors. The symmetry of horizon is
not very important since we need only to keep it invariant at higher
order. It turns out that the asymptotic diffeomorphism vectors forms
a Witt algebra and the corresponding charges of these vectors form
a Virasoro algebra with central extension. Then by using Cardy formula
from CFT, we can derive the entropy of IHs. We should note that such boundary conditions are well satisfied but not proposed explicitly in previous work. These boundary conditions are necessary which make the definition of asymptotic diffeomorphism vectors more precisely.

Our paper is organized as follows. We first review the general framework
of MP's method in section 2. In section 3, the near horizon metric
of IH in Bondi-like coordinate is introduced. The boundary conditions
for the diffeomorphisms related to the Noether current is imposed
in section 4. The asymptotic diffeomorphism vectors are also derived.
In section 5, we calculate the corresponding entropy.  We give the summary and
discussions in section 6.

\section{Noether current from the surface term}

For the sake of completeness, we briefly review the general framework
of the method based on the surface term of gravitational action \cite{Mahji-Padm,Majhi-GH surface,Majhi-EH action,zhang-wang}
in this section. The Gibbons-Hawking term in Einstein gravity is
\begin{eqnarray}
A_{surf} & = & \frac{1}{8\pi G}\int_{\partial M}\sqrt{\gamma}d^{3}xK=\frac{1}{8\pi G}\int_{M}\sqrt{g}d^{4}x\nabla_{a}(KN^{a}).\label{eq:surface-action}
\end{eqnarray}
 Here $\gamma_{ab}$ and $K=-\nabla_{a}N^{a}$ are the induced metric
and the trace of the extrinsic curvature of boundary $\partial M$
of the region $M$, respectively. $N^{a}$ is the unit normal vector
of $\partial M$. Under a general diffeomorphism transformation $x^{a}\rightarrow x^{a}+\xi^{a}$,
the Lagrangian density changes by
\begin{eqnarray}
\delta_{\xi}(\sqrt{g}L) & \equiv & \mathcal{L}_{\xi}(\sqrt{g}L)=\sqrt{g}\nabla_{a}(L\xi^{a}),\label{eq:Dev-L1}
\end{eqnarray}
 where $L=\frac{1}{8\pi G}\nabla_{a}(KN^{a})$. This is a total derivative
so that the action has only a surface contribution. For convenience,
we take an abbreviation $A^{a}\equiv\frac{1}{8\pi G}KN^{a}$ and then
$L=\nabla_{a}A^{a}$. We have
\begin{eqnarray}
\delta_{\xi}(\sqrt{g}\nabla_{a}A^{a}) & \equiv & \mathcal{L}_{\xi}(\sqrt{g}\nabla_{a}A^{a})=\sqrt{g}\nabla_{a}[\nabla_{b}(A^{a}\xi^{b})-A^{b}\nabla_{b}\xi^{a}].\label{eq:Dev-L2}
\end{eqnarray}
 Equating (\ref{eq:Dev-L1}) and (\ref{eq:Dev-L2}), we get the conserved
Noether current
\begin{equation}
J^{a}[\xi]=\nabla_{b}J^{ab}[\xi]=\frac{1}{8\pi G}\nabla_{b}[K(\xi^{a}N^{b}-\xi^{b}N^{a})].\label{eq:curent}
\end{equation}
 The corresponding charge is defined as
\begin{eqnarray}
Q[\xi] & = & \frac{1}{2}\int_{\Sigma}\sqrt{h}d\Sigma_{ab}J^{ab}.\label{eq:Charge}
\end{eqnarray}
 Here $d\Sigma_{ab}=-d^{2}x(N_{a}M_{b}-N_{b}M_{a})$ is the surface
element of the 2-dimensional surface $\Sigma$, $h$ is the determinant
of the corresponding induced metric. $N^{a}$ and $M^{a}$ are asymptotic
spacelike and timelike unit normals of $\Sigma$ in the near horizon
limit. The brackets among the charges are defined by \cite{Mahji-Padm}
\begin{eqnarray}
[Q_{1},Q_{2}] & := & \delta_{\xi_{1}}Q[\xi_{2}]-\delta_{\xi_{2}}Q[\xi_{1}]=\int_{\Sigma}\sqrt{h}d\Sigma_{ab}\left[\xi_{2}^{a}J^{b}[\xi_{1}]-\xi_{1}^{a}J^{b}[\xi^{2}]\right].\label{eq:Charge braket}
\end{eqnarray}
 This definition has not used any field equation and thus is off-shell.
In the following section, we will see that it leads to a Virasoro algebra
with central extension. With the help of central charge and Cardy
formula, the entropy of isolated horizon can be derived.

\section{The near horizon geometry of extreme isolated horizons}

In this section, we review the near horizon limit metric of isolated
horizon briefly. One can refer to \cite{NHLM-IH}  for details.
Roughly speaking, IH is a non-expansion light cone $\Delta$ with
almost stationary inner geometry $(h,D)$. Here $h$ is the induced
metric and $D$ the induced derivative operator. The generator $l$
of $\Delta$ is shear free and keeps $h_{ab}$ and $D$, i.e. $\mathcal{L}_{l}h_{ab}\hat{=}\mathcal{L}_{l}D\hat{=}0$
(Here \textquotedblleft{}$\hat{=}$\textquotedblright{} means equality
holds only on the horizon \textgreek{D}).

We introduce the Bondi-like coordinates to describe the near horizon
geometry of IH \cite{Bondi-coord}. In this coordinate system, we have
a complex null tetrad $\{n,l,m,\bar{m}\}$ which could be expanded
as
\begin{eqnarray}
n & = & \partial_{r},\nonumber \\
l & = & \partial_{u}+U\partial_{r}+X\partial_{\zeta}+\bar{X}\partial_{\bar{\zeta}},\nonumber \\
m & = & W\partial_{r}+\xi\partial_{\vartheta}+\zeta\partial_{\bar{\vartheta}},\label{eq:tetrad}\\
\bar{m} & = & \bar{W}\partial_{r}+\bar{\xi}\partial_{\bar{\vartheta}}+\bar{\zeta}\partial_{\vartheta}.\nonumber
\end{eqnarray}
Vectors $l, m, \bar{m}$ span the tangent space to $\Delta$ and the
metric could be written as
\begin{equation}
g^{ab}=-l^{a}n^{b}-n^{a}l^{b}+m^{a}\bar{m}^{b}+m^{b}\bar{m}^{a}.
\end{equation}
 By definition, it is obvious that coefficients
\begin{equation}
U\hat{=}X\hat{=}W\hat{=}0\label{eq:UXW0}
\end{equation}
 Choosing Bondi gauge $\nabla_{n}(n,l,m,\bar{m})=0$, the near horizon
metric of IH in Bondi-like coordinates finally turn out to be
\begin{equation}
g_{\mu\nu}=\left(\begin{array}{ccc}
-2\kappa_{l}r+(h_{ab}f_{2}^{a}f_{2}^{b}-f_{1})r^{2}+\mathcal{O}(r^{3}) & -1 & -h_{ab}f_{2}^{a}r+\mathcal{O}(r^{2})\\
-1 & 0 & 0\\
-h_{ab}f_{2}^{a}r+\mathcal{O}(r^{2}) & 0 & h_{ab}
\end{array}\right)\label{eq:metric}
\end{equation}
 with
\begin{eqnarray}
f_{1} & = & 6|\pi|^{2}+2Re\Psi_{2}-\frac{R}{12}+2\Phi_{11},\nonumber \\
f_{2}^{a} & = & \pi m^{a}+\bar{\pi}\bar{m}^{a},\label{eq:f2a}\\
h_{ab} & = & m_{a}\bar{m}_{b}+m_{b}\bar{m}_{a}.\nonumber
\end{eqnarray}
 Here $h_{ab}$ is the intrinsic metric of section $\hat{\Delta}$
which is a 2-dimensional space section surrounding the black hole.
$\pi=\bar{m}^{a}l^{b}\nabla_{b}n_{a}$, $\Psi_{2}=-C_{abcd}l^{a}m^{b}n^{c}\bar{m}^{d}$
and $\Phi_{11}=S_{ab}(l^{a}n^{b}+m^{a}\bar{m}^{b})/4$ are the standard
notions in Newman-Penrose formalism \cite{NP form}. Here $C_{abcd}$
is the Weyl tensor, $S_{ab}=R_{ab}-Rg_{ab}/4$ is the traceless Ricci
tensor. $\kappa_{l}$ is the surface gravity of hypersurafce $\Delta$.
We will replace $\kappa_{l}$ by an abbreviated notation $\kappa$
in the following part.

From the deviation of the near horizon metric (\ref{eq:metric}),
we know that the parameters in (\ref{eq:f2a}) are all time independent
on the horizon \cite{NHLM-IH}. But the higher-order terms in the metric (\ref{eq:metric})
can be time dependent. For this reason, isolated horizons can be used to describe non-stationary
cases. On the other hand, only the near horizon limit metric is needed
in the MP's method to calculate the entropy of isolated horizon. The high
order of metric has no contribution to the final result. Thus our
result is applicable for non-stationary horizons. We will discuss more about this at the end of this paper.

For later use, we transform to Schwarzchild-like coordinate by a coordinate
transformation
\begin{eqnarray}
dt=du+\frac{dr}{2\kappa r} & , & d\rho=dr.
\end{eqnarray}
 One then get an asymptotic form near the horizon in Schwarzschild-like
coordinates.
\begin{eqnarray}
ds^{2} & = & g_{uu}dt^{2}-(\frac{g_{uu}}{\kappa\rho}+2)dtd\rho+(\frac{g_{uu}}{4\kappa^{2}\rho^{2}}+\frac{1}{\kappa\rho})d\rho^{2}\\
 &  & +2g_{ui}dtdx^{i}-\frac{g_{ui}}{\kappa\rho}d\rho dx^{i}+h_{ij}dx^{i}dx^{j}.\nonumber
\end{eqnarray}
 Here $g_{uu}=-2\kappa\rho+(h_{ab}f_{2}^{a}f_{2}^{b}-f_{1})\rho^{2}+\mathcal{O}(\rho^{3})$,
$g_{ui}=-h_{ab}f_{2}^{a}\rho+\mathcal{O}(\rho^{2})$. The horizon
is located in $r=\rho=0$.

\section{Boundary conditions}

Now we have the near horizon metric of isolated horizon. To derive
the entropy of IHs, the diffeomorphism vector $\xi$ is remained to
be worked out. In the previous work, only $r-t$ plane is considered due
to the symmetry of metric. The other components of $\xi$ is zero.
However, in our case, IH is more complicated since we do not know
the symmetries of  isolated horizon. Thus there is no good reason
why the other components should not be considered.

To be rigorous, we impose the following boundary conditions from the
very beginning inspired by the work of Strominger \cite{Kerr-CFT}.
\begin{eqnarray}
\delta g_{\mu\nu} & = & \left(\begin{array}{ccc}
\mathcal{O}(r) & \mathcal{O}(r) & \mathcal{O}(r)\\
\mathcal{O}(r) & \mathcal{O}(r) & \mathcal{O}(1)\\
\mathcal{O}(r) & \mathcal{O}(1) & \mathcal{O}(r)
\end{array}\right),\label{eq:boundarycondition}
\end{eqnarray}
 where $\delta g_{\mu\nu}$ is the deviation of the full metric from
the background metric $g_{\mu\nu}$, i.e. $\mathcal{L}_{\xi}g_{\mu\nu}=\delta g_{\mu\nu}$.
This boundary condition is physically acceptable since the deviations
are subleading compared to the full metric except $\delta g_{tt}$
and $\delta g_{ti}$ are of the same order as the leading terms in
(\ref{eq:metric}). We come up with this boundary conditions by assuming
the existence of a non-trivial Virasoro algebra. As stressed by Strominger,
the boundary conditions should not be too strong or too weak.
If the boundary conditions are too strong,  all interesting information
is ruled out and a trivial result will be got. If the boundary conditions
are too weak, we are unable to select the effective information from the
full set of diffeomorphism vectors. For example, if we set $\delta g_{ti}\sim\mathcal{O}(r)$
or $\delta g_{ij}\sim\mathcal{O}(r^{2})$, we will not get desired
algebra any more. We have not found other consistent boundary conditions,
though we cannot prove that condition (\ref{eq:boundarycondition})
is unique. In fact,  these boundary conditions are well satisfied in the previous work \cite{Majhi-GH surface,Majhi-EH action,zhang-wang}.

By requiring $\mathcal{L}_{\xi}g_{\mu\nu}\sim\delta g_{\mu\nu}$ and
$\mathcal{L}_{\xi}\delta g_{\mu\nu}\sim\delta g_{\mu\nu}$, the general
diffeomorphism vectors which preserve this boundary conditions have
the form
\begin{eqnarray}
\xi^{u} & = & F(u,x)+r^{2}F_{1}(u,x)+\mathcal{O}(r^{3}),\nonumber \\
\xi^{r} & = & -r\partial_{u}F(u,x)+\mathcal{O}(r^{2}),\\
\xi^{i} & = & r^{2}G_{1}(u,x)+\mathcal{O}(r^{3}).\nonumber
\end{eqnarray}
 Here $F(u,x)$ and $G_{1}(u,x)$ are regular functions on horizon.
The higher order in $\xi$ may be omitted since it has no affection
on the charges in the following calculations.

Transforming to Schwarzchild-like coordinate system, the diffeomorphism
vector becomes
\begin{eqnarray}
\xi^{t} & = & T-\frac{1}{2\kappa}\partial_{t}T+\mathcal{O}(r^{2}),\nonumber \\
\xi^{\rho} & = & -r\partial_{t}T+\mathcal{O}(r^{2}),\label{eq:xi}\\
\xi^{i} & \sim & \mathcal{O}(r^{2}),\nonumber
\end{eqnarray}
 in which $T(\rho,t,x)=F(u,x)$.

Omitting $\mathcal{O}(r^{2})$, it is obvious that the diffeomorphism
vector $\xi$ and the charge $Q_{\xi}$ are linear in $T$. If we
expand $T$ in terms of a set of basis function $T_{m}$ with
\begin{equation}
T=\sum_{m}A_{m}T_{m},\ \ A_{m}^{*}=A_{-m},
\end{equation}
 we get the corresponding $\xi_{m}^{a}$ and $Q_{m}$ expressed in
the same forms as $T$ replaced by $T_{m}$.

Since there is no symmetry here, the explicit form of $T_{m}$ may
be difficult to be determined. Nonetheless, inspired by previous work,
such as \cite{Carlip2,T-Silva}, we can guess the general form of $T_{m}$
as $T_{m}=\frac{1}{\alpha}exp[im\alpha g(x,t,\rho)]$ where $\alpha$
is an arbitrary constant, $m$ is an integer and $g(x,t,\rho)$ is
a function of the coordinates. The basic function $T_{m}$ should
be orthogonal
\begin{equation}
\int\sqrt{h}d^{2}xT_{m}T_{n}=\frac{A}{\alpha^{2}}\delta_{m+n}\label{eq:ortho}
\end{equation}
 and guarantee the Virasoro algebra

\begin{equation}
[\xi_{m},\xi_{n}]=-i(m-n)\xi_{m+n},\label{eq:Virasoro}
\end{equation}
in which $A$ is the area of the 2-dimensional cross section $\hat{\Delta}$.
The coefficient $\frac{A}{\alpha^{2}}$ in Eq.(\ref{eq:ortho}) is
naturally determined by $\int\sqrt{h}d^{2}x\frac{1}{\alpha^{2}}=\frac{A}{\alpha^{2}}$
when $m+n=0$. Substituting the general form of $T_{m}$ to the commutator
(\ref{eq:Virasoro}), we require
\begin{eqnarray}
\partial_{t}g=1 & , & \partial_{\rho}g=-\frac{1}{2\kappa\rho}.
\end{eqnarray}
Thus the general form of $T_{m}$ which satisfies the orthogonal relation
(\ref{eq:ortho}) and the Virasoro algebra (\ref{eq:Virasoro}) is
\begin{eqnarray}
T_{m} & = & \frac{1}{\alpha}exp[im(\alpha t+\int\frac{-\alpha}{2\kappa\rho}d\rho+P(x))].
\end{eqnarray}
 Here $P(x)$ is a regular function satisfying (\ref{eq:ortho}) on the
horizon. This coincides with the previous work.

\section{Entropy of isolated horizons}

The required diffeomorphism vectors of isolated horizon have been worked
out. Combining Eq.(\ref{eq:curent}, \ref{eq:Charge}, \ref{eq:Charge braket}),
we can calculate the entropy of isolated horizons.

For Einstein gravity, the boundary term is the extrinsic curvature
of the boundary. Taking a hypersurface $\rho=\rho_{c}$, we get a
covector $N_{a}=\frac{(d\rho)_{a}}{\sqrt{g^{\rho\rho}}}=\frac{(d\rho)_{a}}{\sqrt{g^{rr}}}$.
It is asymptotic orthogonal to the horizon $\rho=0$. The trace of
the extrinsic curvature is
\begin{eqnarray}
K=-\nabla_{a}N^{a} & = & -\sqrt{\frac{\kappa}{2r}}+\mathcal{O}(r^{\frac{1}{2}}).
\end{eqnarray}
 $M_{a}$ in Eq.(\ref{eq:Charge}) for the isolated horizon is $M_{a}=-\frac{(dt)_{a}}{\sqrt{-\gamma^{tt}}}=-\sqrt{g^{rr}}(dt)_{a}$.
$N_{a}$ and $M_{a}$ are asymptotic normal to each other $N^{a}M_{a}=1+\frac{g^{rr}}{2\kappa r}\sim\mathcal{O}(r)$
near the horizon.

Plugging (\ref{eq:curent},\ref{eq:xi}) to (\ref{eq:Charge}), the
charge in the near horizon limit $\rho\rightarrow0$ becomes
\begin{eqnarray}
Q_{\xi} & = & \frac{1}{8\pi G}\int\sqrt{h}d^{2}x\left[\kappa T-\frac{1}{2}\partial_{t}T\right].
\end{eqnarray}
 For component $T_{m}$, we get
\begin{equation}
Q_{m}=\frac{A}{8\pi G}\frac{\kappa}{\alpha}\delta_{m,0},
\end{equation}

The commutators of charges lead to
\begin{eqnarray}
[Q_{n},Q_{m}] & = & \frac{1}{8\pi G}\int\sqrt{h}d^{2}x[(\kappa(T_{n}\partial_{t}T_{m}-T_{m}\partial_{t}T_{n})\nonumber \\
 &  & -\frac{1}{2}(T_{n}\partial_{t}^{2}T_{m}-T_{m}\partial_{t}^{2}T_{n})\\
 &  & \frac{1}{4\kappa_{l}}(\partial_{t}T_{n}\partial_{t}^{2}T_{m}-\partial_{t}T_{m}\partial_{t}^{2}T_{n})]\nonumber
\end{eqnarray}
 in the near horizon limit. Substituting the explicit expression of
$T_{m}$, we get
\begin{eqnarray}
[Q_{n},Q_{m}] & = & \frac{i\kappa A}{8\pi G\alpha}(m-n)\delta_{m+n,0}-im^{3}\frac{\alpha A}{16\pi G\kappa}\delta_{m+n,0}.
\end{eqnarray}
 Then the central charge $C$ and zero mode $Q_{0}$ can be read off.
\begin{eqnarray}
\frac{C}{12}=\frac{A}{16\pi G}\frac{\alpha}{\kappa} & , & Q_{0}=\frac{A}{8\pi G}\frac{\kappa}{\alpha}.
\end{eqnarray}

Using the Cardy formula, we get the entropy of IH
\begin{eqnarray}
S & = & 2\pi\sqrt{\frac{CQ_{0}}{6}}=\frac{A}{4G}.
\end{eqnarray}
 This is exactly the Bekenstein-Hawking entropy of IH as expected,
which is consistent with the results in \cite{NHLM-IH,entropy-IH}
from other methods.

\section{Summary and discussions}

The entropy of isolated horizon was derived in MP's approach. Here
we have used the general near horizon limit metric of the IH. No inner symmetry
is involved in our derivation. In previous related works, authors always start from static spherical metric to do calculations. We note that near horizon metric of general spacetimes, stationary or non-stationary, can be put into one unified form. And we demonstrate that only the near horizon metric is needed to derive the entropy in MP's method. So our calculations are not limited to static spherical metrics, but also valid for more general stationary metrics. Moreover, the most significant thing is that our calculations also include the non-stationary cases, for which we do not have an explicit metric in general. Fortunately, we can get the near horizon metric of IHs which have the unified form
as (\ref{eq:metric}). The time dependent behavior of the metric is at high order and turns out to have no influence on the horizon entropy. As a result, all cases appear to have the same entropy formula, the Bekenstein-Hawking entropy formula in Einstein gravity. The result we derived agrees well with the one derived by other methods (thermodynamics, etc).
 Especially, it has been proved that all the axi-symmetric electromagnetic vacuum extreme IHs
coincide with the extreme Kerr-Newman event horizon\cite{KN-IH}.
Thus this method is naturally effective for charged rotating black
holes. For  electromagnetic vacuum
extreme IHs, it coincides with the result in \cite{NHLM-IH}. So our calculations give a general proof that MP's method is universal.

We would like to mention our boundary conditions here. In previous works, such as \cite{Majhi-GH surface,Majhi-EH action,zhang-wang}, they did not give the boundary conditions explicitly due to the symmetries of background. While for genereal IHs, there is no inner symmetry compared to the previous work, so the asymptotic diffeomorphism vector $\xi$ which preserves the boundary condition is more complicated. To be rigorous, we
imposed our boundary conditions near the horizon from the very begining. The asymptotic diffeomorphism
vector respecting the boundary conditions is then calculated. Consistency
requires that $\xi$ be well defined and form a Virasoro algebra.
The boundary conditions we imposed satisfy this requirement. The full
set of the diffeomorphism of the theory is then reduced to a subset
which respects the existence of horizon in a given coordinate system,
and the original gauge degrees of freedom can be thought of as being
effectively upgraded to physical degrees of freedom as far as a particular
class of observers are concerned.

In our calculation, the dimension of spacetime seems to be not important, though calculations will be more complicated.
High dimensional isolated horizons can be considered in future. To
be even more general, one can consider the IHs with comological constant.

\section*{Acknowledgements}

We are very greatful to Prof. Bin Wang for helpful suggestions and discussions. This work is partly supported by the Natural Science Foundation of China under
Grant Nos. 11175245, 11075206.

\end{document}